# Lorentz - Covariant Non - Abelian Gauging of Supermembrane

Hitoshi Nishino[1)] and Subhash Rajpoot[2)]

*Department of Physics & Astronomy*
*California State University*
*1250 Bellflower Boulevard*
*Long Beach, CA 90840*

## Abstract

We perform the Lorentz-covariant non-Abelian gauging of supermembrane (M-2 brane) action. This is a generalization of our previous work based on teleparallel formulation, in which Lorentz covariance was not manifest. We introduce the Killing supervector $\xi^{AI}$ with the adjoint index $I$ for a non-Abelian gauge group $H$. This formulation is applicable to the compactification of supermembrane from eleven dimensions into $D$ dimensions, such as $H = SO(11-D)$ for the compactification $M_{11} \to S^{11-D} \times M_D$ $(1 \le D \le 9)$.

[1)] E-Mail: hnishino@csulb.edu
[2)] E-Mail: rajpoot@csulb.edu

## 1. Introduction

In our previous paper [1], we have performed the gauging of supermembrane action [2] for an arbitrary non-Abelian group with the Killing supervector $\xi^{AI}$, where the index $I$ is for the adjoint representation of the gauge group. However, the drawback of that formulation [1] was the lack of Lorentz covariance. In a sense, this is inevitable, because any compactification from eleven-dimensional (11D) superspace [3] is possible at the expense of the original Lorentz covariance in 11D. Nevertheless, for practical applications it is more advantageous to maintain 'formal' Lorentz covariance as much as possible, in terms of Killing supervectors. The new technique needed is certain Lorentz-covariant compactification in superspace.

Compactifications formulated in superspace [4], at the expense of Lorentz covariance are not necessarily new. For example in [5], simple dimensional reductions were performed from 10D superspace into 4D superspace, consistently as Green-Schwarz superstring background [6]. However, in such simple dimensional reductions, some field variables in superspace lose their degrees of freedom, and moreover, the Lorentz covariance in the original 10D superspace becomes less manifest. In the Lorentz-covariant gauging of M-2 brane, we can compensate such drawback by formal Lorentz covariance by the use of Killing supervectors. As a matter of fact, such a formulation for an Abelian case [7] has been recently provided from the viewpoint of super-embedding [8].

In view of such developments, it is the next natural step to perform a Lorentz-covariant formulation for non-Abelian gauge groups that are applicable to general compactifications from 11D. Once a Lorentz-covariant non-Abelian generalization is established, such a formulation will provide a powerful methodology to investigate M-theory [9], in particular, its compactifications into lower dimensions.

Motivated by these observations, in our present paper, we will generalize our previous result [1] to Lorentz-covariant formulation, instead of relying on teleparallel superspace which is too restrictive. In other words, we will generalize the gauge group from $U(1)$ [7][8] to a general non-Abelian gauge group $H$ which arises as the isotropy group for the $(11-D)$-dimensional internal compact manifold $B_{11-D} = G/H$ in the compactification $M_{11} \rightarrow B_{11-D} \times M_D$. A typical example is $H = SO(7)$ for $B_7 = S^7 = SO(8)/SO(7)$ [10].

In the next section, we will prepare certain geometrical relationships needed for our gauging. In section 3, we will show the invariances of our action, in particular, not only fermionic $\kappa$ transformations [11] and local-Lorentz transformations under certain restrictions, but also the peculiar composite $\Lambda$ transformation that are originally associated with the third-rank superpotential $B_{ABC}$. We will see that this transformation is modified in the presence of the Killing supervector $\xi^{AI}$ for the non-Abelian group $H$. Compared with the Abelian case [7][8], our system is further generalized to non-Abelian group $H$ applicable to a general compactification $M_{11} \to B_{11-D} \times M_D$, where $B_{11-D}$ has the coset structure $B_{11-D} = G/H$.

## 2. Preliminaries about Modified Superspace with Killing Supervectors

We first arrange the basic relationships for the Lorentz covariant formulation of non-Abelian gauging of supermembrane [2]. The most fundamental quantity is the Killing supervector $\xi^{AI}$ satisfying [1]

$$E_A \xi^{BI} = \xi^{CI} C_{CA}{}^B \ , \tag{2.1}$$

where $E_A \equiv E_A{}^M \partial_M$ and the $C$'s is the superspace anholonomy coefficient $C_{AB}{}^C \equiv (E_{[A} E_{B)}{}^M) E_M{}^C$ [4]. We use the indices $A \equiv (a,\alpha)$, $B \equiv (b,\beta)$, $\cdots$ for the local coordinates in the 11D target superspace, where $a, b, \cdots = (0), (1), \cdots, (10)$ are local bosonic coordinates, while $\alpha, \beta, \cdots = 1, 2, \cdots, 32$ are for the local fermionic coordinates. Accordingly, our 11D metric is $(\eta_{ab}) = \text{diag.} (+, -, -, \cdots, -)$. The index $I$ is for the adjoint representation of the gauged non-Abelian group $H$. Eq. (2.1) is now casted into Lorentz-covariant form

$$\nabla_A \xi^{BI} = \xi^{CI} T_{CA}{}^B + U_A{}^{BI} \ , \quad U_A{}^{BI} \equiv \xi^{CI} \omega_{CA}{}^B \ . \tag{2.2}$$

At first glance, $U_A{}^{BI}$ looks like a Lorentz-non-covariant superfield. However, it is easy to show that this tensor is Lorentz-covariant as in [12], as well as by the Lorentz transformation rule (3.11d) given later.

For our supermembrane action, we need the gauged version of the so-called pull-back in the target 11D superspace [2][1]

$$\Pi_i{}^A \equiv \left( \partial_i Z^M - m A_i{}^I \xi^{MI} \right) E_M{}^A \ . \tag{2.3}$$

In this paper, we use the indices $i, j, \cdots = 0, 1, 2$ for the curved 3D worldvolume coordinates with the metric $(\eta_{ij}) = \text{diag.} (+, -, -)$. For the curved coordinates in the 11D superspace, we use $M \equiv (m,\mu)$, $N \equiv (n,\nu)$, $\cdots$, where $m, n, \cdots = 0, 1, \cdots, 10$ (or $\mu, \nu, \cdots = 1, 2, \cdots, 32$) are for the curved bosonic (or fermionic) coordinates. In (2.3), $m$ is a coupling constant with the dimension of mass, while $A_i{}^I$ is the 3D non-Abelian vector field, whose field strength is[3)]

$$F_{ij}{}^I \equiv \partial_{[i} A_{j]}{}^I + m f^{IJK} A_i{}^J A_j{}^K \ , \tag{2.4}$$

with the structure constant $f^{IJK}$ of the gauge group $H$. Compared with the conventional supermembrane [2], the new ingredient is the minimal coupling of the non-Abelian vector field $A_i{}^I$ on the 3D worldvolume [1].

As usual in 11D superspace [3], we define the superfield strength $G_{ABCD}$ for the third-rank superpotential $B_{ABC}$, as well as the supertorsion and supercurvature:

$$\begin{aligned} G_{ABCD} &\equiv +\tfrac{1}{3!} \nabla_{[A} B_{CDE)} - \tfrac{1}{2 \cdot 2} T_{[AB|}{}^F B_{F|CD)} \ , \\ T_{AB}{}^C &\equiv +\tfrac{1}{2} (\nabla_{[A} E_{B)}{}^M) E_M{}^C = (E_{[A} E_{B)}{}^M) E_M{}^C - \omega_{[AB)}{}^C \ , \\ R_{AB}{}^{cd} &\equiv + E_{[A} \omega_{B)}{}^{cd} - C_{AB}{}^E \omega_E{}^{cd} - \omega_{[A|}{}^{ce} \omega_{|B) e}{}^d \end{aligned} \tag{2.5}$$

where our Lorentz covariant derivative $\nabla_M$ acts on an arbitrary supervector $V_B$ as

$$\begin{aligned} \nabla_M V_B &\equiv \partial_M V_B - \tfrac{1}{2} \omega_M{}^{de} \mathcal{M}_{de} \circ V_B \equiv \partial_M V_B - \tfrac{1}{2} \omega_M{}^{de} (\mathcal{M}_{de})_B{}^C V_C \\ &\equiv \partial_M V_B - \omega_{MB}{}^C V_C \ . \end{aligned} \tag{2.6}$$

Here $\mathcal{M}$ is the Lorentz connection generator acting as

$$(\mathcal{M}^{ab})_C{}^D \equiv \begin{cases} (\mathcal{M}^{ab})_c{}^d \equiv +\delta_c{}^{[a} \eta^{b]d} \ , \\ (\mathcal{M}^{ab})_\gamma{}^\delta \equiv +\tfrac{1}{2} (\gamma^{ab})_\gamma{}^\delta \ , \end{cases} \tag{2.7}$$

so that we have conveniently

$$\tfrac{1}{2} \omega_A{}^{de} (\mathcal{M}_{de})_B{}^C \equiv \omega_{AB}{}^C \ , \quad \tfrac{1}{2} R_{AB}{}^{ef} (\mathcal{M}_{ef})_C{}^D \equiv R_{ABC}{}^D \ . \tag{2.8}$$

The Bianchi identities (BIds) for $T_{AB}{}^C$, $G_{ABCD}$ and $R_{AB}{}^{CD}$ are the usual ones:

$$\tfrac{1}{2} \nabla_{[A} T_{BC)}{}^D - \tfrac{1}{2} T_{[AB|}{}^E T_{E|B)}{}^D + \tfrac{1}{2} R_{[ABC)}{}^D \equiv 0 \ , \tag{2.9a}$$

$$\tfrac{1}{4!} \nabla_{[A} G_{BCDE)} - \tfrac{1}{2(3!)} T_{[AB|}{}^F G_{F|CDE)} \equiv 0 \ , \tag{2.9b}$$

$$\tfrac{1}{2} \nabla_{[A} R_{BC)}{}^{DE} - \tfrac{1}{2} T_{[AB|}{}^F R_{F|C)}{}^{DE} \equiv 0 \ . \tag{2.9c}$$

---

[3)] For the symbol of (anti)symmetrization, we put *no* normalization factor, e.g., $X_{[AB)} \equiv X_{AB} - (-1)^{AB} X_{BA}$ without $1/2$.

We frequently use the *tilded* superfield for an arbitrary totally antisymmetric supertensor $X_{A_1 A_2 \cdots A_n}$ in this paper, defined generally by

$$\widetilde{X}_{A_1 A_2 \cdots A_{n-1}}{}^I \equiv \xi^{BI} X_{BA_1 A_2 \cdots A_{n-1}} \quad . \tag{2.10}$$

Typical examples are[4]

$$\widetilde{G}_{ABC}{}^I \equiv \xi^{DI} G_{DABC} \quad , \quad \widetilde{\check{G}}_{ABC}{}^I \equiv \xi^{DI} \check{G}_{DABC} \quad , \quad \widetilde{\check{\check{G}}}_{AB}{}^{IJ} \equiv \xi^{CI} \widetilde{\check{G}}_{CAB}{}^J \quad , \tag{2.11a}$$

$$\widetilde{\Lambda}_A{}^I \equiv \xi^{BI} \Lambda_{BA} \quad , \quad \widetilde{\widetilde{\Lambda}}{}^{IJ} \equiv \xi^{AI} \widetilde{\Lambda}_A{}^J \quad . \tag{2.11b}$$

Interestingly enough, $\widetilde{G}_{ABC}{}^I$ can be expressed in terms of $\widetilde{B}_{AB} \equiv \xi^C B_{CAB}$, as

$$\widetilde{G}_{ABC}{}^I \equiv \xi^{DI} G_{DABC} = - \left( \tfrac{1}{2} \nabla_{[A} \widetilde{B}_{BC)}{}^I - \tfrac{1}{2} T_{[AB|}{}^D \widetilde{B}_{D|C)}{}^I \right) \quad , \tag{2.12}$$

as if $\widetilde{B}_{AB}{}^I$ were the potential superfield for the superfield strength $\widetilde{G}_{ABC}{}^I$. Note the extra overall negative sign needed in the r.h.s.

Due to the existence of the Killing supervector $\xi^{MI}$, we have the following Lie-derivative relationships[5]

$$\mathcal{L}_{\xi^I} \xi^{MJ} \equiv +\xi^{P[I|} \partial_P \xi^{M|J]} = +f^{IJK} \xi^{MK} = \xi^{BI} \xi^{CJ} C_{CB}{}^M \quad , \tag{2.13a}$$

$$\mathcal{L}_{\xi^I} E_M{}^A \equiv +\xi^{NI} \partial_N E_M{}^A + (\partial_M \xi^{NI}) E_N{}^A = 0 \quad , \tag{2.13b}$$

$$\mathcal{L}_{\xi^I} E_A{}^M \equiv +\xi^{NI} \partial_N E_A{}^M - E_A \xi^{MI} = 0 \quad . \tag{2.13c}$$

Similarly, the following superpotential or superfield strengths have zero Lie derivatives:

$$\mathcal{L}_{\xi^I} B_{ABC} \equiv \xi^{DI} E_D B_{ABC} = 0 \quad , \tag{2.14a}$$

$$\mathcal{L}_{\xi^I} C_{AB}{}^C \equiv \xi^{DI} E_D C_{AB}{}^C = 0 \quad , \quad \mathcal{L}_{\xi^I} T_{AB}{}^C \equiv \xi^{DI} E_D T_{AB}{}^C = 0 \quad , \tag{2.14b}$$

$$\mathcal{L}_{\xi^I} \omega_A{}^{BC} \equiv \xi^{DI} E_D \omega_A{}^{BC} = 0 \quad , \quad \mathcal{L}_{\xi^I} R_{AB}{}^{CD} \equiv \xi^{DI} E_D R_{AB}{}^{CD} = 0 \quad . \tag{2.14c}$$

Due to (2.13b,c), it does not matter whether the local Lorentz indices $A, B, \cdots$ or curved indices $M, N, \cdots$ are used for Lie derivatives, *e.g.*,

$$\mathcal{L}_{\xi^I} B_{MNP} \equiv \xi^{QI} \partial_Q B_{MNP} + \tfrac{1}{2} (\partial_{[M|} \xi^{QI}) B_{Q|NP)} = 0 \quad , \tag{2.15}$$

and the like.

---

[4] The $\check{G}_{ABCD}$ will be defined shortly in (2.16c).

[5] For the transformation properties of Killing vectors, see *e.g.*, [13].

In our system, we also need the *checked* superfield strengths defined by

$$\check{C}_{AB}{}^C \equiv C_{AB}{}^C - m^3 \widetilde{B}_{AB}{}^I \xi^{CI} \ , \quad \check{T}_{AB}{}^C \equiv T_{AB}{}^C - m^3 \widetilde{B}_{AB}{}^I \xi^{CI} \ , \quad (2.16a)$$

$$\check{R}_{AB}{}^{CD} \equiv R_{AB}{}^{CD} + m^3 \widetilde{B}_{AB}{}^I U^{CDI} \ , \quad (2.16b)$$

$$\check{G}_{ABCD} \equiv G_{ABCD} + \tfrac{1}{2\cdot2\cdot2} m^3 \widetilde{B}_{[AB}{}^I \widetilde{B}_{CD)}{}^I \ . \quad (2.16c)$$

These naturally arise in the formulation with Killing supervectors [1][8]. They satisfy the modified BIds

$$\tfrac{1}{2}\nabla_{[A}\check{T}_{BC)}{}^D - \tfrac{1}{2}\check{T}_{[AB|}{}^E \check{T}_{E|C)}{}^D + \tfrac{1}{2}\check{R}_{[ABC)}{}^D - m^3 \widetilde{\check{G}}_{ABC}{}^I \xi^{DI} \equiv 0 \ , \quad (2.17a)$$

$$\tfrac{1}{4!}\nabla_{[A}\check{G}_{BCDE)} - \tfrac{1}{2(3!)}\check{T}_{[AB|}{}^F \check{G}_{F|CDE)} \equiv 0 \ , \quad (2.17b)$$

$$\tfrac{1}{2}\nabla_{[A}\check{R}_{BC)}{}^{DE} - \tfrac{1}{2}\check{T}_{[AB|}{}^F \check{R}_{F|C)}{}^{DE} + m^3 \widetilde{\check{G}}_{ABC}{}^I U^{DEI} \equiv 0 \ . \quad (2.17c)$$

Note the *absence* of explicit $m$-dependent terms in the $\check{G}$-BId. The Abelian case of these BIds coincides with those given in [8]. The third power of $m$ in (2.17a,c) can be understood by the mass dimensions, *e.g.*, $[\widetilde{\check{G}}_{abc}{}^I] = 0$, $[\xi^{aI}] = -1$ and $[U^{deI}] = 0$.

The superspace constraints are in terms of these *checked* supertorsions, supercurvatures and supertensors. They are listed here at the mass dimensions $d \leq 2$, as

$$\check{T}_{\alpha\beta}{}^c = +i(\gamma^c)_{\alpha\beta} \ , \quad \check{T}_{\alpha b}{}^c = 0 \ , \quad \check{T}_{\alpha\beta}{}^\gamma = 0 \ , \quad \check{T}_{ab}{}^c = 0 \ , \quad (2.18a)$$

$$\check{G}_{\alpha\beta cd} = +\tfrac{1}{2}(\gamma_{cd})_{\alpha\beta} \ , \quad \check{G}_{\alpha\beta\gamma\delta} = 0 \ , \quad \check{G}_{\alpha\beta\gamma d} = 0 \ , \quad (2.18b)$$

$$\check{T}_{\alpha b}{}^\gamma = +\tfrac{i}{144}(\gamma_b{}^{cdef}\check{G}_{cdef} + 8\gamma^{cde}\check{G}_{bcde})_\alpha{}^\gamma$$
$$- \tfrac{i}{4}m^3(\gamma_d)_\alpha{}^\gamma \xi_b{}^I \xi^{dI} + \tfrac{i}{24}m^3(\gamma_b)_\alpha{}^\gamma (\xi^{dI})^2 \ , \quad (2.18c)$$

$$\check{R}_{\alpha\beta cd} = +\tfrac{1}{72}(\gamma_{cd}{}^{efgh})_{\alpha\beta}\check{G}_{efgh} + \tfrac{1}{3}(\gamma^{ef})_{\alpha\beta}\check{G}_{cdef}$$
$$- \tfrac{1}{2}m^3(\gamma_{[c|}{}^e)_{\alpha\beta}\xi_{|d]}{}^I \xi_e{}^I + \tfrac{1}{12}m^3(\gamma_{cd})_{\alpha\beta}(\xi^{eI})^2 \ , \quad (2.18d)$$

$$\check{R}_{\alpha bcd} = -\tfrac{i}{2}(\gamma_{[c}\check{T}_{d]b})_\alpha + \tfrac{i}{2}(\gamma_b\check{T}_{cd})_\alpha - \tfrac{1}{4}m^3(\gamma_{b[c|}\xi^I)_\alpha \xi_{|d]}{}^I + \tfrac{1}{4}m^3(\gamma_{cd}\xi^I)_\alpha \xi_b{}^I \ , \quad (2.18e)$$

$$\nabla_\gamma \check{T}_{ab}{}^\delta = -\tfrac{1}{4}(\gamma^{cd})_\gamma{}^\delta \check{R}_{abcd} - \tfrac{i}{144}(\gamma_{[a|}{}^{[4]})_\gamma{}^\delta \nabla_{|b]}\check{G}_{[4]} + \tfrac{i}{18}(\gamma^{[3]})_\gamma{}^\delta \nabla_{[a}\check{G}_{b][3]}$$
$$- \tfrac{1}{2}m^3(\gamma_{ab}\xi^I)_\gamma \xi^{\delta I} - \tfrac{i}{4}m^3(\gamma^e)_\gamma{}^\delta \nabla_{[a}(\xi_{b]}{}^I \xi_e{}^I) - \tfrac{i}{24}m^3(\gamma_{[a|})_\gamma{}^\delta \nabla_{|b]}[(\xi^{cI})^2]$$
$$- \tfrac{1}{(144)^2}(\gamma_{[a|}{}^{[4]}\gamma_{|b]}{}^{[4]'})_\gamma{}^\delta \check{G}_{[4]}\check{G}_{[4]'} - \tfrac{1}{(18)^2}(\gamma^{[3]}\gamma^{[3]'})_\gamma{}^\delta \check{G}_{[a|[3]}\check{G}_{|b][3]'}$$
$$+ \tfrac{1}{144\cdot18}[\gamma^{[3]},\gamma_{[a|}{}^{[4]}]_\gamma{}^\delta \check{G}_{|b][3]}\check{G}_{[4]} + \tfrac{1}{576}m^3[\gamma_{[a|}{}^{[4]},\gamma_c]_\gamma{}^\delta \check{G}_{[4]}\xi_{|b]}{}^I \xi^{cI}$$

$$- \tfrac{1}{3456}m^3[\gamma_{[a}{}^{[4]},\gamma_{b]}]_\gamma{}^\delta \check{G}_{[4]}(\xi^{cI})^2 + \tfrac{1}{72}m^3[\gamma^{[3]},\gamma_c]_\gamma{}^\delta \check{G}_{[a|[3]}\xi_{|b]}{}^I \xi^{cI}$$
$$+ \tfrac{1}{432}m^3[\gamma^{[3]},\gamma_{[a|}]_\gamma{}^\delta \check{G}_{|b][3]}(\xi^{cI})^2 - \tfrac{1}{8}m^6(\gamma^{fg})_\gamma{}^\delta(\xi_a{}^I\xi_b{}^J)(\xi_f{}^I\xi_g{}^J)$$
$$- \tfrac{1}{288}m^6(\gamma_{ab})_\gamma{}^\delta[(\xi^{cI})^2]^2 - \tfrac{1}{48}m^6(\gamma_{c[a|})_\gamma{}^\delta \xi_{|b]}{}^I \xi^{cI}(\xi^{dJ})^2 \quad , \tag{2.18f}$$

$$\nabla_\alpha \check{G}_{bcde} = -\tfrac{1}{8}(\gamma_{[bc}\check{T}_{de]})_\alpha \quad , \tag{2.18g}$$

where $(\gamma_{ab}\xi^I)_\gamma \equiv (\gamma_{ab})_\gamma{}^\delta \xi_\delta{}^I$ or $(\gamma_{[c}\check{T}_{d]b})_\alpha \equiv -(\gamma_{[c|})_{\alpha\delta}\check{T}_{|d]b}{}^\delta$, etc. in order to save space. We also use the subscripts $[n]$ for totally antisymmetric vectorial $n$ indices, e.g., $(\gamma^{[3]})_\gamma{}^\delta \nabla_{[a}\check{G}_{b][3]} \equiv (\gamma^{cde})_\gamma{}^\delta \nabla_{[a}\check{G}_{b]cde}$. The high powers in $m$ in (2.18c) through (2.18f) can be understood from $[\xi^{aI}] = -1$. The comparison of our results with [8] is easily done, e.g. the relative ratio between the two terms at $\mathcal{O}(m^3)$ in (2.18c) is in agreement with eq. (7.11) of [8].

The pull-back $\Pi_i{}^A$ satisfies its proper Bianchi identity

$$\partial_{[i}\Pi_{j]}{}^A \equiv \Pi_i{}^C \Pi_j{}^B C_{BC}{}^A - mF_{ij}{}^I\xi^{AI} \quad , \tag{2.19a}$$

$$\nabla_{[i}\Pi_{j]}{}^A \equiv \Pi_i{}^C \Pi_j{}^B T_{BC}{}^A - mF_{ij}{}^I\xi^{AI} \quad , \tag{2.19b}$$

where $\nabla_{[i}\Pi_{j]}{}^A \equiv \partial_{[i}\Pi_{j]}{}^A - \Pi_{[i|}{}^B\omega_B{}^{AC}\Pi_{|j]C}$.

The $\Gamma$'s is defined by the product of three $\gamma$-matrices in 11D, as usual [2]. It satisfies various relationships that are valid only by use of the $g_{ij}$-field equation

$$g_{ij} \doteq \eta_{ab}\Pi_i{}^a\Pi_j{}^b \quad . \tag{2.20}$$

where $\doteq$ implies a field equation, distinguished from an algebraic equation. Relevant relationships are

$$\Gamma \equiv + \tfrac{i}{6\sqrt{g}}\epsilon^{ijk}\Pi_i{}^a\Pi_j{}^b\Pi_k{}^c\gamma_{abc} \equiv +\tfrac{i}{6\sqrt{g}}\epsilon^{ijk}\gamma_{ijk} \quad , \tag{2.21a}$$

$$\Gamma^2 \doteq +I \quad , \tag{2.21b}$$

$$\gamma_i \doteq +\tfrac{i}{2\sqrt{g}}\epsilon^{ijk}\gamma_{jk}\Gamma \quad , \tag{2.21c}$$

$$\gamma_i \equiv +\Pi_i{}^a\gamma_a \quad , \quad \gamma_{ij} \equiv \Pi_i{}^a\Pi_j{}^b\gamma_{ab} \quad . \tag{2.21d}$$

These are formally the same as in the non-gauged case [2], the Abelian case [8], or our previous Lorentz-non-covariant formulation [1].

## 3. Supermembrane Action and Invariances

Prepared with the fundamental geometric relationships at hand, we are ready to consider the action for our non-Abelian gauged supermembrane:

$$I \equiv \int d^3\sigma \, \mathcal{L} = \int d^3\sigma \, \Big[ \, +\tfrac{1}{2}\sqrt{g}\, g^{ij}\, \eta_{ab}\, \Pi_i{}^a \Pi_j{}^b - \tfrac{1}{2}\sqrt{g} + \tfrac{1}{3}\epsilon^{ijk}\, \Pi_i{}^C \Pi_j{}^B \Pi_k{}^A B_{ABC}$$
$$+ \tfrac{1}{2} m^{-1} \epsilon^{ijk} \left( F_{ij}{}^I A_k{}^I - \tfrac{1}{3} m f^{IJK} A_i{}^I A_j{}^J A_k{}^K \right) \Big] \, . \tag{3.1}$$

Due to the 3D metric $(\eta_{ij}) \equiv \text{diag.}(+,-,-)$ engaged, we need no negative sign in $\sqrt{g}$. In this paper, we assign the mass dimensions

$$[m] = +1 \, , \quad [A_i{}^I] = [B_{abc}] = [g_{ij}] = 0 \, , \quad [F_{ij}{}^I] = +1 \, ,$$
$$[\xi^{aI}] = -1 \, , \quad [\xi^{\alpha I}] = -1/2 \, , \quad [\Pi_i{}^a] = 0 \, , \quad [\Pi_i{}^\alpha] = +1/2 \, . \tag{3.2}$$

so that we need the negative power $m^{-1}$ in the Chern-Simons (CS) term in (3.1). Even though the first line in (3.1) looks exactly the same as in the conventional supermembrane action [2], there is a minimal coupling involved in the $\Pi$'s.

Our action is invariant under the fermionic $\kappa$ transformation rule

$$\delta_\kappa E^\alpha = + (I + \Gamma)^{\alpha\beta} \kappa_\beta \equiv + [(I + \Gamma)\kappa]^\alpha \, , \quad \delta_\kappa E^a = 0 \, , \tag{3.3a}$$

$$\delta_\kappa E_A{}^M = + (\delta_\kappa E^B) E_B E_A{}^M \, , \quad \delta_\kappa E_M{}^A = + (\delta_\kappa E^B) E_B E_M{}^A \, , \tag{3.3b}$$

$$\delta_\kappa A_i{}^I = + m^2 \Pi_i{}^A \xi^{BI} (\delta_\kappa E^C) B_{CBA} \equiv + m^2 \Pi_i{}^A \xi^{BI} \Lambda_{BA} \equiv + m^2 \Pi_i{}^A \widetilde{\Lambda}_A{}^I \equiv + m^2 \widetilde{\Lambda}_i{}^I \, , \tag{3.3c}$$

$$\delta_\kappa \xi^{AI} = - \xi^{CI} (\delta_\kappa E^B) C_{BC}{}^A \, , \tag{3.3d}$$

$$\delta_\kappa \Pi_i{}^A = + \partial_i (\delta_\kappa E^A) + \Pi_i{}^C (\delta_\kappa E^B) \check{C}_{BC}{}^A + m^3 \widetilde{\Lambda}_i{}^I \xi^{AI}$$
$$= + \nabla_i (\delta_\kappa E^A) + \Pi_i{}^C (\delta_\kappa E^B)(\check{T}_{BC}{}^A + \omega_{BC}{}^A) + m^3 \widetilde{\Lambda}_i{}^I \xi^{AI} \, , \tag{3.3e}$$

$$\delta_\kappa B_{ABC} = + (\delta_\kappa E^D) E_D B_{ABC} \, . \tag{3.3f}$$

In (3.3c) and (3.3e), we used the $\Lambda$'s defined by $\Lambda_{AB} \equiv (\delta_\kappa E^C) B_{CAB}$, $\widetilde{\Lambda}_A{}^I \equiv \xi^{BI} \Lambda_{BA}$ and $\widetilde{\Lambda}_i{}^I \equiv \Pi_i{}^A \widetilde{\Lambda}_A{}^I$. Even though we will use the same symbols such as $\Lambda_{AB}$ for the $\delta_\Lambda$ transformation (3.9), the $\Lambda$'s used here are different.

The confirmation of $\delta_\kappa I = 0$ is quite parallel to the usual case without the gauging with the coupling constant $m$. The only difference is that now all the superspace constraints are in terms of *checked* superfields $\check{T}_{AB}{}^C$, $\check{R}_{AB}{}^{CD}$ and $\check{G}_{ABCD}$ in (2.16). Also as usual, we need to use the relationships in (2.20) and (2.21), while the variation $(\delta_\kappa g_{ij})(\delta I / \delta g_{ij})$ is to be understood as the first-order formalism, as long as the algebraic $g_{ij}$-field equation (2.20) holds [2].

One remark is in order. In the evaluation of $\delta_\kappa I$, we need to form a superfield strength $\check{G}_{ABCD}$. Here we need the subtle relationship

$$\check{G}_{ABCD} \equiv +\tfrac{1}{3!}\nabla_{[A}B_{BCD)} - \tfrac{1}{2\cdot 2}T_{[AB|}{}^E B_{E|CD)} + \tfrac{1}{2\cdot 2\cdot 2}m^3 \widetilde{B}_{[AB}{}^I \widetilde{B}_{CD)}{}^I$$

$$= +\tfrac{1}{3!}\nabla_{[A}B_{BCD)} - \tfrac{1}{2\cdot 2}\check{T}_{[AB|}{}^E B_{E|CD)} - \tfrac{1}{2\cdot 2\cdot 2}m^3 \widetilde{B}_{[AB}{}^I \widetilde{B}_{CD)}{}^I \ . \tag{3.4}$$

Note the sign flip in the $m^3$-term between the first and second expressions, caused by the *checked* supertorsion in the second term.

Our action is also invariant under local non-Abelian transformation for the group $H$:

$$\delta_\alpha A_i{}^I = +\partial_i \alpha^I + m f^{IJK} A_i{}^J \alpha^K \equiv D_i \alpha^I \ , \tag{3.5a}$$

$$\delta_\alpha Z^M = +m\alpha^I \xi^{MI} \ , \tag{3.5b}$$

$$\delta_\alpha \xi^{MI} = +m\alpha^J \xi^{NJ} \partial_N \xi^{MI} \ , \quad \delta_\alpha \xi^{AI} = -m f^{IJK} \alpha^J \xi^{AK} \ , \tag{3.5c}$$

$$\delta_\alpha E_A{}^M = +m\alpha^I E_A \xi^{MI} = +m\alpha^I \xi^{NI} \partial_N E_A{}^M \ , \tag{3.5d}$$

$$\delta_\alpha E_M{}^A = -m\alpha^I (\partial_M \xi^{NI}) E_N{}^A = +m\alpha^I \xi^{NI} \partial_N E_M{}^A \ , \tag{3.5e}$$

$$\delta_\alpha \Pi_i{}^A = 0 \ , \quad \delta_\alpha g_{ij} = 0 \ , \quad \delta_\alpha B_{ABC} = 0 \ , \quad \delta_\alpha \omega_A{}^{BC} = 0 \ , \tag{3.5f}$$

$$\delta_\alpha \widetilde{B}_{AB}{}^I = -m f^{IJK} \alpha^J \widetilde{B}_{AB}{}^K \ , \quad \delta_\alpha U_{AB}{}^I = -m f^{IJK} \alpha^J U_{AB}{}^K \ , \tag{3.5g}$$

The $\delta_\alpha$-invariance $\delta_\alpha I = 0$ is straightforward to confirm by the use of (3.5f).

There is a very important aspect associated with $\delta_\alpha$ transformation and BIds. Note that the superfield strengths $G_{ABCD}$, $T_{AB}{}^C$, $R_{AB}{}^{CD}$ as well as their *checked* ones $\check{G}_{ABCD}$, $\check{T}_{AB}{}^C$, $\check{R}_{AB}{}^{CD}$ are *not* invariant under $\delta_\alpha$. However, it is not too difficult to confirm that the BIds (2.9) or (2.17) are invariant under the $\delta_\alpha$ transformations (3.5). The easiest way is to consider first the following BIds in terms of curved indices

$$\tfrac{1}{2}\nabla_{[M}T_{NP)}{}^A + \tfrac{1}{2}R_{[MNP)}{}^A \equiv 0 \ , \tag{3.6a}$$

$$\tfrac{1}{4!}\partial_{[M}G_{NPQR)} \equiv 0 \ , \tag{3.6b}$$

$$\tfrac{1}{2}\nabla_{[M}R_{NP)}{}^{AB} \equiv 0 \ , \tag{3.6c}$$

which are equivalent to (2.9). Next use the relationships

$$\delta_\alpha B_{MNP} = -\tfrac{1}{2}m\alpha^I (\partial_{[M|}\xi^{QI}) B_{Q|NP)} \ , \tag{3.7a}$$

$$\delta_\alpha G_{MNPQ} = + \tfrac{1}{2}(-1)^{NR} m\alpha^I (\partial_M \xi^R)(\partial_N B_{RPQ}) + \text{(23 more terms)} \ , \tag{3.7b}$$

$$\delta_\alpha T_{MN}{}^A = -(-1)^{NP} m\alpha^I (\partial_{[M} \xi^{PI})(\partial_{N)} E_P{}^A) + m\alpha^I (\partial_{[M|} \xi^{PI}) \omega_{P|N)}{}^A$$
$$- (-1)^{NP} m\alpha^I (\partial_{[M} \xi^{PI}) \omega_{N)P}{}^A \ , \tag{3.7c}$$

$$\delta_\alpha \omega_M{}^{AB} = - m\alpha^I (\partial_M \xi^{NI}) \omega_N{}^{AB} \ , \tag{3.7d}$$

$$\delta_\alpha R_{MN}{}^{AB} = (-1)^{NP} m\alpha^I (\partial_M \xi^{PI}) \left[ \partial_N \omega_P{}^{AB} + (-1)^{N(C+A+P)} \omega_P{}^{[A|C} \omega_{NC}{}^{|B)} \right]$$
$$+ \text{(one more term)} \ , \tag{3.7e}$$

which are just the corollaries of (3.5). In (3.7b), '23 more terms' are needed for the total antisymmetrization of $[MNPQ)$ with their appropriate Grassmann parities. Similarly for (3.7e), we need 'one more term', in such a way that the total expression on the r.h.s. is (anti)symmetric in $M \leftrightarrow N$. The most important technique is as follows. For example, even though $\delta_\alpha G_{MNPQ}$ itself is non-zero, this does *not* disturb the $G$-BId (3.6b), due to the exact-form structure of $\delta_\alpha G_{MNPQ}$ in (3.7b):

$$\delta_\alpha \left[ \tfrac{1}{4!} \partial_{[M} G_{NPQR)} \right] = 0 \ . \tag{3.8}$$

We can confirm that other BIds in (3.6) are also consistent with $\delta_\alpha$ transformations.

Another important feature is that the difference between the *checked* and *non-checked* superfield strengths does not affect the consistency of all the BIds and $\delta_\alpha$ transformation. The reason is that the difference terms, such as $\widetilde{B}_{AB}{}^I \xi^{CI}$, $\widetilde{B}_{AB}{}^I U_C{}^{DI}$ or $\widetilde{B}_{[AB}{}^I \widetilde{B}_{CD)}{}^I$ are all invariant under $\delta_\alpha$, as can be easily seen from (3.5g). The $\delta_\alpha$-invariance of the superspace BIds also implies that all the target superspace superfield equations are consistent with $\delta_\alpha$ transformations.

This situation is in a sense similar to the global $E_{8(+8)}$ symmetry of $N=8$ supergravity in 4D [14] where this global symmetry is realized only at the field-equation level, but not at the field-strength level. However, the difference is that our $\delta_\alpha$ symmetry is local symmetry, while the former is global. Another difference is that this $\delta_\alpha$ symmetry is realized as the 3D action invariance of supermembrane.

Our action is also invariant under composite $\Lambda$ transformation, which was associated with the superpotential $B_{MNP}$

$$\delta_\Lambda B_{ABC} = + \tfrac{1}{2} E_{[A} \Lambda_{BC)} - \tfrac{1}{2} C_{[AB|}{}^D \Lambda_{D|C)} = + \tfrac{1}{2} \nabla_{[A} \Lambda_{BC)} - \tfrac{1}{2} T_{[AB|}{}^D \Lambda_{D|C)} \ , \tag{3.9a}$$

$$\delta_\Lambda E_A{}^M = -m^3 \widetilde{\Lambda}_A{}^I \xi^{MI} ~, \qquad \delta_\Lambda E_M{}^A = +m^3 \widetilde{\Lambda}_M{}^I \xi^{AI} ~, \tag{3.9b}$$

$$\delta_\Lambda A_i{}^I = +m^2 \Pi_i{}^A \widetilde{\Lambda}_A{}^I \equiv +m^2 \widetilde{\Lambda}_i{}^I ~, \tag{3.9c}$$

$$\delta_\Lambda \xi^{AI} = +m^3 \xi^{BI} \widetilde{\Lambda}_B{}^J \xi^{AJ} \equiv +m^3 \widetilde{\widetilde{\Lambda}}^{IJ} \xi^{AJ} ~, \tag{3.9d}$$

$$\delta_\Lambda \Pi_i{}^A = 0 ~, \quad \delta_\Lambda \omega_A{}^{BC} = 0 ~, \quad \delta_\Lambda g_{ij} = 0 ~, \quad \delta_\Lambda Z^M = 0 ~, \quad \delta_\Lambda \xi^{MI} = 0 ~, \tag{3.9e}$$

$$\delta_\Lambda G_{ABCD} = +\tfrac{1}{2} m^3 \left( E_{[A} \widetilde{\Lambda}_{B)}{}^I - \tfrac{1}{2} C_{[AB|}{}^E \widetilde{\Lambda}_E{}^I \right) \widetilde{B}_{|CD)}{}^I ~, \tag{3.9f}$$

$$\delta_\Lambda \widetilde{B}_{AB}{}^I = -E_{[A} \widetilde{\Lambda}_{B)}{}^I + C_{AB}{}^C \widetilde{\Lambda}_C{}^I + m^3 \widetilde{\widetilde{\Lambda}}^{IJ} \widetilde{B}_{AB}{}^J$$

$$= -\nabla_{[A} \widetilde{\Lambda}_{B)}{}^I + T_{AB}{}^C \widetilde{\Lambda}_C{}^I + m^3 \widetilde{\widetilde{\Lambda}}^{IJ} \widetilde{B}_{AB}{}^J ~, \tag{3.9g}$$

$$\delta_\Lambda C_{AB}{}^C = -m^3 \left( E_{[A} \widetilde{\Lambda}_{B)}{}^I - C_{AB}{}^D \widetilde{\Lambda}_D{}^I \right) \xi^{CI} ~, \tag{3.9h}$$

$$\delta_\Lambda U_{AB}{}^I = +m^3 \widetilde{\widetilde{\Lambda}}^{IJ} U_{AB}{}^J ~, \tag{3.9i}$$

$$\delta_\Lambda R_{ABC}{}^D = +m^3 \left( E_{[A} \widetilde{\Lambda}_{B)}{}^I - C_{AB}{}^E \widetilde{\Lambda}_E{}^I \right) U_C{}^{DI} ~, \tag{3.9j}$$

$$\delta_\Lambda \widetilde{\widetilde{G}}_{ABC}{}^I = +m^3 \widetilde{\widetilde{\Lambda}}^{IJ} \widetilde{\widetilde{G}}_{ABC}{}^J ~, \tag{3.9k}$$

$$\delta_\Lambda \check{C}_{AB}{}^C = 0 ~, \quad \delta_\Lambda \check{T}_{AB}{}^C = 0 ~, \quad \delta_\Lambda \check{G}_{ABCD} = 0 ~, \quad \delta_\Lambda \check{R}_{ABC}{}^D = 0 ~. \tag{3.9\ell}$$

As has been mentioned, the $\Lambda_{AB}$ here is *not* the one used in (3.3) for $\delta_\kappa$ transformation. The composite infinitesimal parameter superfield $\Lambda_{AB} = \Lambda_{AB}(Z^M)$ in (3.9) is arbitrary, *except for* the Lie-derivative constraint

$$\mathcal{L}_{\xi^I} \Lambda_{AB} \equiv \xi^{MI} \partial_M \Lambda_{AB} = \xi^{CI} E_C \Lambda_{AB} = 0 ~. \tag{3.10}$$

As (3.9$\ell$) shows, all these *checked* superfield strengths are invariant under $\delta_\Lambda$. The non-trivial $\delta_\Lambda$ transformation necessitates the $m$-dependent modification of the superfield strengths as in (2.16). Relevantly, all the BIds (2.17) for the *checked* superfield strengths are consistent with $\delta_\Lambda$ transformations, including also $\xi^{DI}$, $\widetilde{\widetilde{G}}_{ABC}{}^I$ and $U^{DEI}$. The higher powers of $m$ in (3.9) can be understood in terms of mass dimensions, e.g., $[\Pi_i{}^a] = 0$, $[\Lambda_{ab}] = -1$, $[\widetilde{\Lambda}_a{}^I] = -2$, $[\widetilde{\widetilde{\Lambda}}^{IJ}] = -3$. The confirmation of the invariance $\delta_\Lambda I = 0$ is straightforward under (3.9a,e) and (2.19a) with the aid of (3.10).

Finally and most importantly, our action has the local Lorentz invariance $\delta_\lambda I = 0$ with the parameter superfield $\lambda^{AB} = \lambda^{AB}(Z^M) = -(-1)^{AB} \lambda^{BA}$:

$$\delta_\lambda \omega_M{}^{AB} = \partial_M \lambda^{AB} - \omega_M{}^{[A|C} \lambda_C{}^{|B)} \equiv \nabla_M \lambda^{AB} ~, \tag{3.11a}$$

$$\delta_\lambda E_A{}^M = +\lambda_A{}^B E_B{}^M ~, \qquad \delta_\lambda E_M{}^A = -E_M{}^B \lambda_B{}^A ~, \tag{3.11b}$$

$$\delta_\lambda \xi^{MI} = 0 , \quad \delta_\lambda \xi^{AI} = \lambda^{AB} \xi_B{}^I , \quad \delta_\lambda \Pi_i{}^M = 0 , \quad \delta_\lambda \Pi_i{}^A = \lambda^{AB} \Pi_{iB} , \quad \delta_\lambda g_{ij} = 0 , \quad (3.11c)$$

$$\delta_\lambda B_{ABC} = +\tfrac{1}{2}\lambda_{[A|}{}^D B_{D|BC)} , \quad \delta_\lambda U_{AB}{}^I = +\lambda_{[A|}{}^C U_{C|B)}{}^I , \quad (3.11d)$$

and similarly for other Lorentz-covariant supertensors, such as $T_{AB}{}^C$, $\check{T}_{AB}{}^C$, $G_{ABCD}$, $\check{G}_{ABCD}$, $R_{AB}{}^{CD}$ and $\check{R}_{AB}{}^{CD}$, *etc.* that we do not write explicitly here. Most importantly, $\lambda_{AB}$ undergoes the Lie-derivative constraint

$$\mathcal{L}_{\xi^I} \lambda_{AB} \equiv \xi^{MI} \partial_M \lambda_{AB} = \xi^{CI} E_C \lambda_{AB} = 0 . \quad (3.12)$$

This condition guarantees the Lorentz covariance of $U_{AB}{}^I$ in (3.11d), as has been promised in section 2. All of our BIds are also manifestly locally Lorentz covariant, including the new $m$-dependent terms. Under (3.11c,d) our action is manifestly locally Lorentz invariant: $\delta_\lambda I = 0$.

Note that the coefficient $m^{-1}$ in front of the CS term in (3.1) is to be quantized [15] for most of non-Abelian groups whose $\pi_3$-homotopy mappings are non-trivial [16]:

$$\pi_3(H) = \begin{cases} \mathbb{Z} & (\text{for } H \neq U(1), SO(2), SO(4), Spin(4)) \\ \mathbb{Z} \oplus \mathbb{Z} & (\text{for } H = SO(4)) \end{cases} , \quad (3.13)$$

## 4. Superfield Equations

As we have fixed the constraints (2.18) at mass dimensions $d \leq 2$, we are ready to get superfield equations at $d \geq 3/2$. The first one is the gravitino superfield equation at $d = 3/2$:

$$+ i(\gamma^c)_{\gamma\delta} \check{T}_{ac}{}^\delta - \tfrac{1}{6} m^3 (\gamma_{ac})_{\gamma\delta} \xi^{\delta I} \xi^{cI} - \tfrac{4}{3} m^3 \xi_\gamma{}^I \xi_a{}^I \doteq 0 . \quad (4.1)$$

As usual, this is obtained from the $T$-BI starting $\nabla_{(\beta} \check{T}_{\gamma)a}{}^\delta - \cdots \equiv 0$ at $d = 3/2$, by contracting its $\gamma$ and $\delta$ indices. As mentioned before, the third power of $m$ is due to the mass dimensions $[\xi^{aI}] = -1$, $[\xi^{\alpha I}] = -1/2$. As usual, the gravitational superfield equation at $d = 2$ can be obtained by the operation $i(\gamma_b)^{\gamma\epsilon} \nabla_\epsilon$ on (4.1):[6]

$$\check{R}_{ab} - \tfrac{1}{3} \check{G}_{a[3]} \check{G}_b{}^{[3]} + \tfrac{1}{36} \eta_{ab} (\check{G}_{[4]})^2 - \tfrac{1}{3} m^3 \eta_{ab} (\overline{\xi}^I \xi^I)$$
$$+ m^6 \Big[ \tfrac{1}{2} (\xi_a{}^I \xi_b{}^J)(\xi_c{}^I \xi^{cJ}) - \tfrac{1}{4} (\xi_a{}^I \xi_b{}^I)(\xi_c{}^J)^2 - \tfrac{1}{12} \eta_{ab} (\xi_c{}^I \xi^{cJ})^2 + \tfrac{1}{24} \eta_{ab} \{(\xi^{cI})^2\}^2 \Big] \doteq 0 , \quad (4.2)$$

---

[6] Due to the Lorentz connection convention (2.6), the relative sign between the Ricci tensor and the $\check{G}^2$-term is opposite to the conventional case.

where $(\overline{\xi}^I \xi^I) \equiv \xi^{\alpha I} \xi_\alpha{}^I$. Eq. (4.2) in turn yields the scalar curvature superfield equation

$$\check{R} - \tfrac{1}{36}(\check{G}_{[4]})^2 - \tfrac{5}{24}m^6 \left[2(\xi_a{}^I \xi^{aJ})^2 - \{(\xi_a{}^I)^2\}^2\right] - \tfrac{11}{3}m^3(\overline{\xi}^I \xi^I) \doteq 0 \ . \tag{4.3}$$

Similarly, the $B_{abc}$-superfield equation is obtained by multiplying (4.1) by $(\gamma_{de})^{\alpha\gamma}\nabla_\alpha$ and antisymmetrizing the indices $[ade]$, or alternatively, by multiplying (4.1) by $(\gamma^{def}\gamma^a)^{\alpha\gamma}\nabla_\alpha$. Both methods give consistently the same result

$$\nabla_d \check{G}_{abc}{}^d + \tfrac{1}{576}\epsilon_{abc}{}^{[4][4]'}\check{G}_{[4]}\check{G}_{[4]'} - \tfrac{1}{4}m^3 U_{[ab}{}^I \xi_{c]}{}^I + \tfrac{i}{2}m^3(\overline{\xi}^I \gamma_{abc} \xi^I) \doteq 0 \ , \tag{4.4}$$

where $(\overline{\xi}^I \gamma_{abc} \xi^I) \equiv \xi^{\alpha I}(\gamma_{abc})_\alpha{}^\beta \xi_\beta{}^I$. Compared with (4.3), there is no $m^6 \xi^4$ term present in (4.4).

In the derivation of the $B_{abc}$-superfield equation, we need the relationships, such as

$$\nabla_A \xi^{BI} = \xi^{CI}\check{T}_{CA}{}^B - m^3 \widetilde{\widetilde{B}}_A{}^{IJ} \xi^{BJ} + U_A{}^{BI} \ , \tag{4.5}$$

with $\widetilde{\widetilde{B}}_A{}^{IJ} \equiv \xi^{BI}\widetilde{B}_{BA}{}^J \equiv \xi^{BI}\xi^{CJ}B_{CBA}$. Even though both sides of (4.5) are *not* covariant under the $\delta_\Lambda$ transformation, we can easily see that $\widetilde{\widetilde{B}}_A{}^{IJ}$ plays a role as a 'connection'. In fact, (4.5) is equivalent to

$$\mathcal{D}_A \xi^{BI} \equiv \nabla_A \xi^{BI} + m^3 \widetilde{\widetilde{B}}_A{}^{IJ}\xi^{BJ} = \xi^{CI}\check{T}_{CA}{}^B + U_A{}^{BI} \ , \tag{4.6}$$

where the r.h.s. is manifestly $\delta_\Lambda$-covariant under (3.9d,i,$\ell$). In this sense, $\mathcal{D}_A$ is a $\delta_\Lambda$-covariant derivative. The explicit components of (4.6) within our constraints (2.18) are

$$\mathcal{D}_\alpha \xi^{bI} = +i(\gamma^b)_{\alpha\gamma} \xi^{\gamma I} \ , \quad \mathcal{D}_a \xi^{bI} = +U_a{}^{bI} \ , \tag{4.7a}$$

$$\mathcal{D}_a \xi^{\beta I} = +\xi^{\gamma I}\check{T}_{\gamma a}{}^\beta \ , \quad \mathcal{D}_\alpha \xi^{\beta I} = -\xi^{dI}\check{T}_{\alpha d}{}^\beta + U_\alpha{}^{\beta I} \ . \tag{4.7b}$$

For example, due to $T_{\alpha b}{}^c = 0$ in (2.18a), there is no $\xi \check{T}$ term in $\mathcal{D}_a \xi^{bI}$. Eq. (4.7a) also yields the familiar relationship

$$\mathcal{D}_{(a} \xi_{b)}{}^I = 0 \ , \tag{4.8}$$

in a $\delta_\Lambda$-covariant fashion.

Eq. (4.6) further implies that

$$[\mathcal{D}_A, \mathcal{D}_B\} \xi^{CI} = -\check{R}_{AB}{}^{CD}\xi_D{}^I + \check{T}_{AB}{}^D \mathcal{D}_D \xi^{CI} - m^3 \widetilde{\widetilde{G}}_{AB}{}^{IJ}\xi^{CJ} \ , \tag{4.9}$$

where $\widetilde{\widetilde{\check{G}}}_{AB}{}^{IJ} \equiv \xi^{CI}\widetilde{\check{G}}_{CAB}{}^{J}$. Eq. (4.9) is very natural in terms of *checked* covariant derivatives and superfield strengths. This provides other evidence of the 'covariance' of the covariant derivative $\mathcal{D}_A$, and total consistency of the system in terms of $\mathcal{D}_A$.

## 5. Concluding Remarks

We have in this paper performed the locally Lorentz-covariant non-Abelian gauging of M-2 brane [2]. We have confirmed the four invariances of our action: the fermionic invariance $\delta_\kappa I = 0$, the non-Abelian gauge invariance $\delta_\alpha I = 0$, the composite $\Lambda$-invariance $\delta_\Lambda I = 0$, and the most important local Lorentz invariance $\delta_\lambda I = 0$. We have shown that the BIds should be modified by the *checked* superfield strengths $\check{T}_{AB}{}^C$, $\check{R}_{AB}{}^{CD}$ and $\check{G}_{ABCD}$ together with the $m$-dependent terms, as in (2.17). As a special case, we have seen that the Abelian version agrees with the result in [8].

As a technical development, we have noticed that our $\delta_\alpha$ symmetry of our supermembrane action is not the symmetry of the target superspace superfield strengths $G_{ABCD}$, $T_{AB}{}^C$ or $R_{AB}{}^{CD}$. Even though they are *not* invariant under the $\delta_\alpha$ transformation, all the BIds are consistent with the $\delta_\alpha$ transformation. This situation resembles the global $E_{8(+8)}$ symmetry realized only at the field-equation level, but not at the field-strength or lagrangian level, in $N = 8$ supergravity in 4D [14]. The differences, however, are (i) Our $\delta_\alpha$ symmetry is local symmetry, while the former [14] is global. (ii) Our $\delta_\alpha$ symmetry is realized as the 3D action invariance of supermembrane.

We have also derived all the superfield equations, such as the gravitino superfield equation (4.1), the gravitational superfield equation (4.2), and the $B_{abc}$-superfield equation (4.4). In all of these superfield equations, we have seen the peculiar involvement of the Killing supervectors $\xi^{AI}$. In particular, in the $B_{abc}$-superfield equation, we have seen the presence of $U^{bcI} \equiv \xi^{AI}\omega_A{}^{bc}$. To our knowledge, we have not encountered in the past these superfield equations with *non-Abelian* Killing supervectors in the target superspace superfield equations with the particular combination $U_{BC}{}^I \equiv \xi^{AI}\omega_{ABC}$.

In section 4, we have also seen the total consistency of our system formulated in terms of the $\delta_\Lambda$-covariant derivative $\mathcal{D}_A$. In particular, the commutator on the Killing supervector $\xi^{CI}$ (4.9) shows the closure of the gauge algebra in terms of $\mathcal{D}_A$.

Once we have established the non-Abelian gauging of M-2 brane, we have many applications to compactifications into lower dimensions from 11D. The simplest case is from 11D into 10D with $H=U(1)$, as studied also in [8]. Next less non-trivial case is from 11D into 9D on a sphere: $M_{11} \to S^2 \times M_9$ where $S^2 \approx SO(3)/SO(2)$ and $H = SO(2)$. In the general case of $M_{11} \to S^{11-D} \times M_D$ with a round sphere $S^{11-D}$ for $1 \leq D \leq 8$, we can identify $H = SO(11-D)$, since $S^{11-D} \approx SO(12-D)/SO(11-D)$. Other examples for compactifications into 4D are summarized as

| $B_7 = G/H$ | $G$ | $H$ | Refs. |
|---|---|---|---|
| Round $S^7$ | $SO(8)$ | $SO(7)$ | [10] |
| Squashed $S^7$ | $Sp(2) \times Sp(1)$ | $Sp(1) \times Sp(1)$ | [17] |
| $M^{pqr}$ | $SU(3) \times SU(2) \times U(1)$ | $SU(2) \times U(1) \times U(1)$ | [18] |
| $N^{pqr}$ | $SU(3) \times U(1)$ | $U(1) \times U(1)$ | [19] |
| $Q^{pqr}$ | $SU(2) \times SU(2) \times SU(2)$ | $U(1) \times U(1)$ | [20] |

Table 1: Examples of $B_7 = G/H$ for Compactification $M_{11} \to B_7 \times M_4$

The non-Abelian gauging of supermembrane necessitates the existence of the worldvolume gauge field $A_i{}^I$ with the CS term, because the minimal coupling of $A_i{}^I$ in the superspace pullback $\Pi_i{}^A$ *via* the Killing supervector $\xi^{AI}$ necessitates a CS term in the M-2 brane action, required by local fermionic invariance [11].

As long as the gauge group $H$ in $B_{11-D} = G/H$ is compact, *e.g.*, $H = SO(11-D)$, it is likely that $H$ has the non-trivial $\pi_3$-homotopy mapping (3.13) [16]. This implies that the coefficient $m^{-1}$ for the CS term should be quantized [15]. Since the constant $m$ controls all the new couplings, including the target superfield equations, such a quantization affects all the $m$-dependent coefficients of these superfield equations. This aspect was not clearly understood in the conventional Kaluza-Klein formulation, providing another non-trivial consequence of our formulation.

We believe that the formulation, methodology and the results in this paper will open a new avenue for investigating M-theory [9], M-2 brane [2], or other extended objects.

This work is supported in part by NSF Grant # 0652996.